\documentclass[a4paper,english,aps,twocolumn]{revtex4}
\usepackage[T1]{fontenc}
\usepackage[latin1]{inputenc}
\usepackage{babel}
\usepackage{graphics}

\makeatletter

\providecommand{\LyX}{L\kern-.1667em\lower.25em\hbox{Y}\kern-.125emX\@}

\makeatother
\begin{document}

\title{The dependence of cosmological parameters estimated from the microwave
background on non-gaussianity}

\author{Luca Amendola }

\affiliation{Osservatorio Astronomico di Roma, Via Frascati, 00040 Monteporzio
Catone, Italy}

\date{\today {}}

\begin{abstract}
The estimation of cosmological parameters from cosmic microwave experiments
has almost always been performed assuming gaussian data. In this paper
the sensitivity of the parameter estimation to different assumptions
on the probability distribution of the fluctuations is tested. Specifically,
adopting the Edgeworth expansion, I show how the cosmological parameters
depend on the skewness of the \( C_{\ell } \) spectrum. In the particular
case of skewness independent of \( \ell  \) I find that the primordial
slope, the baryon density and the cosmological constant increase with
the skewness.
\end{abstract}
\maketitle

\section{Introduction}

The new generation of cosmic microwave background (CMB) experiments
(see e.g. \cite{lee, halverson, net}, and the future missions Planck
\cite{planck}, and Map \cite{map}) promises to estimate the cosmological
parameters within a precision of 1\%. The current dataset already
allows in some cases an uncertainty below 10\% on such parameters
as the baryon density or the primordial spectral slope. Such a precision
allows and demands a clear assessment of the theoretical assumptions.

So far, all the estimations of cosmological parameters based on the
CMB data assumed a Gaussian distribution of the primordial fluctuations
(the only exception I know of is Ref. \cite{cont} in which the primordial
slope \( n \) was estimated in presence of skewness). Such an assumption
is based on the conventional models of inflation and has the obvious
and enormous advantage of being simple and uniquely determined. However,
the gaussianity of the primordial temperature fluctuations is still
to be fully tested \cite{ban, kogut, mag} and there exist several
theoretical models which actually predict its violation \cite{falk}.
Therefore, it is necessary to quantify how the cosmological parameters
derived from CMB experiments depend on the statistical properties
of the fluctuation field.

In this paper I derive the dependence of four cosmological parameters
on the skewness of the fluctuations assuming a flat space. The parameters
are the primordial slope \( n \), the baryon and cold dark matter
rescaled density parameters \( \omega _{b}\equiv \Omega _{b}h^{2},\omega _{c}\equiv \Omega _{c}h^{2} \)
( \( h \) is the Hubble constant in units of 100 km/sec/Mpc), and
the cosmological constant density parameter \( \Omega _{\Lambda } \).
The flat space constraint reduces to the relation \( h^{2}=(\omega _{b}+\omega _{c})/(1-\Omega _{\Lambda }). \) 

It is clear that removing the hypothesis of Gaussianity leaves room
for an infinity of different possible assumptions concerning the fluctuation
distribution. I adopt here the Edgeworth expansion (EE), for three
reasons: a) it can be seen as a perturbation of a Gaussian function;
b) it is easy to manipulate analytically and c) it is the distribution
followed by any random variable that is a linear combination of \( N \)
random variables in the limit of large \( N \) (for \( N\rightarrow \infty  \)
the Edgeworth distribution reduces to a Gaussian). The latter property
might be useful to describe fluctuations that arise due to several
independent sources. The EE has been previously used in cosmology
to model small deviations from Gaussianity \cite{ame94, ame96, ju, fos, ameastro}.

The main drawback of the EE is that it is not positive definite. However,
when the deviation from Gaussianity is small, this problem is pushed
many standard deviations away from the peak and does not affect the
parameter estimation.

This paper is meant to exemplify the effects that a non-zero skewness
introduces on the likelihood estimation. For generality, I will not
confine myself to any specific mechanism for generating the non-gaussianity.
Moreover, for simplicity, I will skip over several additional complications
like bin cross-correlations, calibration, pointing and beam errors
that an accurate analysis should take into account.

\section{Edgeworth likelihood}

The likelihood function usually adopted in CMB studies (e.g. \cite{bon, lee, halverson, net})
is an offset log-normal function. This function is an approximation
to the exact likelihood that holds for Gaussian data in presence of
Gaussian noise \cite{bon}. The offset depends on quantities that
are not yet publicly available; since the offset can be neglected
in the limit of small noise, we assume as starting point a simple
log-normal that, neglecting factors independent of the variables,
can be written as

\begin{equation}
-2\log L(\alpha _{j})=\sum _{i}\frac{Z_{\ell ,t}(\ell _{i};\alpha _{j})-Z_{\ell ,d}(\ell _{i})}{\sigma _{\ell }^{2}},
\end{equation}
 where \( Z_{\ell }\equiv \log \hat{C}_{\ell } \), the subscripts
\( t \) and \( d \) refer to the theoretical quantity and to the
real data, \( \hat{C}_{\ell } \) are the spectra binned over some
interval of multipoles centered on \( \ell _{i} \), \( \sigma _{\ell } \)
are the experimental errors on \( Z_{\ell ,d} \), and the parameters
are denoted collectively as \( \alpha _{j} \). We neglect also the
residual correlation between multipole bins, which should be anyway
very small for the latest data. An overall amplitude parameter \( A \)
can be integrated out analytically adopting a logarithmic measure
\( d\log A \) in the likelihood. Writing \( \hat{C}_{\ell }=A\hat{C}_{\ell }^{\prime } \)
it follows \( Z_{\ell ,t}=\log A+\log \hat{C}_{\ell ,t}^{\prime }=\log A+Z_{\ell ,t}^{\prime } \)
so that, neglecting the factors independent of the variables and putting
\( w=\log A \), we obtain\begin{equation}
L\propto \int \exp \left[ -\sum _{\ell }\frac{\left[ w+\Delta _{\ell }\right] ^{2}}{2\sigma _{\ell }^{2}}\right] dw\propto e^{-\frac{1}{2}\left( \gamma -\frac{\beta ^{2}}{\alpha }\right) },
\end{equation}
 where 

\begin{eqnarray*}
\Delta _{\ell } & = & Z'_{\ell ,t}(\ell _{i};\alpha _{j})-Z'_{\ell ,d}(\ell _{i}),\\
\alpha  & = & \sum 1/\sigma _{\ell }^{2},\quad \beta =\sum \Delta _{\ell }/\sigma _{\ell }^{2}.\\
\gamma  & = & \sum \Delta _{\ell }^{2}/\sigma _{\ell }^{2}.
\end{eqnarray*}

Let us now introduce the Edgeworth expansion. Denoting with \( x_{\ell }=\left( w+\Delta _{\ell }\right) /\sigma _{\ell } \)
the normal variable in the Gaussian function, the Edgeworth expansion
is \cite{kendall}\begin{eqnarray}
P(x_{\ell })=\exp \left[ -\frac{x^{2}_{\ell }}{2}\right] \left[ \right. 1+\frac{k_{3,\ell }}{6}H_{3}(x_{\ell })+ &  & \nonumber \\
\frac{k_{4,\ell }}{24}H_{4}(x_{\ell })+\frac{k_{3,\ell }^{2}}{72}H_{6}(x_{\ell }) & +...,\left. \right] , & 
\end{eqnarray}
where \( k_{n,\ell } \) is the \( n \)-th cumulant of \( x_{\ell } \)
and \( H_{n} \) is the Hermite polynomial of \( n \)-th order. Notice
that the EE has the same norm, mean and variance as the Gaussian,
but different mode (the peak of the distribution). Here, as a first
step, we limit ourselves to the first non-Gaussian term containing
the skewness \( k_{3,\ell } \).

Assuming that \( x_{\ell } \) is distributed according to the Edgeworth
expansion, we can build the truncated Edgeworth likelihood function
\cite{ameastro} to first order in \( k_{3,\ell } \): \[
L=e^{-\sum _{\ell }\frac{x^{2}_{\ell }}{2}}\left[ 1+\frac{\kappa _{3,\ell }}{6}\sum _{\ell }H_{3}(x_{\ell })\right] ,\]
 with \( H_{3}(x_{\ell })=x_{\ell }^{3}-3x_{\ell } \). Now, integrating
over \( w \) we obtain\begin{eqnarray}
L=\int e^{-\sum \frac{x^{2}_{\ell }}{2}}\left[ 1+\frac{1}{6}\sum k_{3,\ell }H_{3}(x_{\ell })\right] dw= &  & \nonumber \\
\sqrt{\frac{2\pi }{\alpha }}e^{-\frac{1}{2}\left( \gamma -\frac{\beta ^{2}}{\alpha }\right) }\left[ 1+\frac{1}{6}g(k_{3,\ell },\Delta _{\ell },\sigma _{\ell })\right] , &  & \label{edgelik} 
\end{eqnarray}
where\begin{eqnarray*}
 &  & \\
\alpha ^{3}g(k_{3,\ell },\Delta _{\ell },\sigma _{\ell })=(-3\alpha \beta -\beta ^{3})\sum \frac{k_{3,\ell }}{\sigma _{\ell }^{3}} &  & \\
+(3\alpha ^{2}+3\alpha \beta ^{2})\sum \frac{k_{3,\ell }\Delta _{\ell }}{\sigma _{\ell }^{3}}-3\alpha ^{2}\beta \sum k_{3,\ell }\left( \frac{\Delta ^{2}_{\ell }}{\sigma _{\ell }^{3}}-\frac{1}{\sigma _{\ell }}\right)  &  & \\
+\alpha ^{3}\sum k_{3,\ell }\left( \frac{\Delta ^{3}_{\ell }}{\sigma _{\ell }^{3}}-3\frac{\Delta _{\ell }}{\sigma _{\ell }}\right) . &  & \label{fung} \\
 & 
\end{eqnarray*}
 This is the likelihood function that we study below. The effect of
the extra terms is to shift the peak (or mode) of the distribution
of each \( C_{\ell } \) while leaving the mean unperturbed. Since
the shift depends on \( \Delta _{\ell } \) ,\( \sigma _{\ell } \)
and \( k_{3,\ell } \), the resulting \emph{mode spectrum} will be
distorted with respect to the \emph{mean spectrum}. Therefore, the
likelihood maximization will produce in general results that depend
on \( k_{3,\ell } \). In Fig. 1 we show the peak shift introduced
in the simplified case in which the skewness is independent of \( \ell  \):
if \( k_{3} \) is \emph{negative}, the spectrum is shifted \emph{upward}
by a larger amount at the very small and very large multipoles, and
by a smaller amount around \( \ell =200 \), where the relative errors
are the smallest; if \( k_{3} \) is positive the shift is downward.
As a consequence of the distortion, we expect that a constant negative
skewness favours spectra which are tilted downward with respect to
the Gaussian case, and the contrary for a positive skewness. In general,
the cosmological parameters will depend on the multipole dependence
of \( k_{3,\ell } \). For small \( k_{3}\sigma _{\ell } \), the
shift can be approximated by\begin{equation}
C_{\ell \, (mode)}=C_{\ell (mean)}(1-k_{3}\sigma _{\ell }/2).
\end{equation}
Clearly, if the peak shift introduced by the EE were independent of
\( \ell  \), the integration over the amplitude \( w \) would erase
the non-gaussian effect on the likelihood. That is, putting \( \sigma _{\ell } \)
and \( \Delta _{\ell } \) equal to a constant independent of \( \ell  \)
we obtain \( g(k_{3,\ell },\Delta _{\ell },\sigma _{\ell })=0. \)

\begin{figure}
{\centering \resizebox*{10cm}{!}{\includegraphics{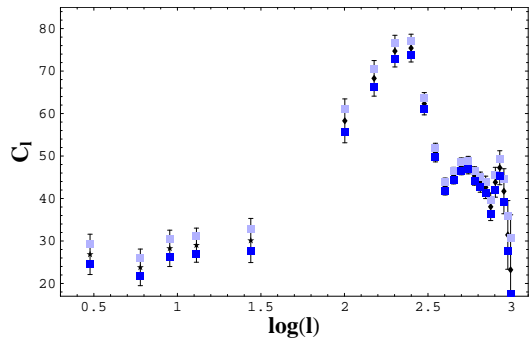}} \par}

\caption{The squares mark the mode of the \protect\( C_{\ell }\protect \)
distribution for \protect\( k_{3}=-2\protect \) (light squares),
and for \protect\( k_{3}=2\protect \) (dark squares). The data are
from Cobe and Boomerang.}
\end{figure}

\section{Dependence on the skewness}

To evaluate the likelihood, a library of CMB spectra is generated
using CMBFAST \cite{sel}. Following \cite{net} I adopt the following
uniform priors: \( n\in (0.7,1.3), \) \( \quad \omega _{b}\in (0.0025,0.08),\quad  \)\ \( \omega _{c}\in (0.05,0.4),\Omega _{\Lambda }\in (0,0.9) \). As
extra priors, the value of \( h \) is confined in the range \( (0.45,0.9) \)
and the universe age is limited to \( >10 \) Gyr. The remaining input
parameters requested by the CMBFAST code are set as follows: \( T_{cmb}=2.726K, \)
\( Y_{He}=0.24,N_{\nu }=3.04,\tau _{c}=0. \) In the analysis of \cite{net}
\( \tau _{c} \), the optical depth to Thomson scattering, was also
included in the general likelihood and, in the flat case, was found
to be compatible with zero at slightly more than 1\( \sigma  \) .
Therefore here, to reduce the parameter space, I assume \( \tau _{c} \)
to vanish. The theoretical spectra are compared to the data from COBE
\cite{bon} and Boomerang \cite{net}.

To specify the skewness \( k_{3,\ell } \) three simplified cases
are studied: in the first one ({}``constant skewness{}''), \( k_{3,\ell }=k_{3}^{*} \)
is assumed independent of the multipole \( \ell  \); in the second
({}``gaussian skewness{}''), the skewness is assumed to be generated
by some process only in a particular range of multipoles:\begin{equation}
k_{3,\ell }=k_{3}^{*}e^{-(\ell -\ell ^{*})^{2}/2\Delta ^{*2}_{\ell }},
\end{equation}
where, in the numerical examples below, I put \( \ell ^{*}=200 \)
and \( \Delta _{\ell }^{*}=20 \). In the third case ({}``hierarchical
skewness{}''), the {}``hierarchical{}'' ansatz is assumed \cite{peeb},
in which the skewness of the temperature field is proportional to
the square of its variance. At the first order, we can assume that
the skewness of the \( C_{\ell } \) distribution is proportional
to the skewness of the fluctuation field, so I put\begin{equation}
\label{hier}
k_{3,\ell }=k_{3}^{*}\left( C_{\ell }/C^{*}\right) ^{2},
\end{equation}
where, for instance, \( C^{*}=C_{200} \). In all three cases \( k^{*}_{3} \)
is left as a free parameter. These three choices are of course purely
an illustration of what a real physical mechanism might possibly produce.

Fig. 2 shows the one-dimensional Edgeworth likelihood functions marginalized
in turn over the other three parameters. For the {}``constant skewness{}'',
\( k_{3}^{*} \) varies from -1.6 to 1.2 ( light to dark curves):
below and above these values the likelihood begins to show pronounced
negative wings, which signals that the first order Edgeworth expansion
is no longer acceptable. While the likelihood for \( \omega _{c} \)
is almost independent of \( k_{3} \) , it turns out that the other
likelihoods move toward higher values for higher skewness. As anticipated,
this can be explained by observing that a higher skewness implies
smaller \( C_{\ell } \) at small multipoles: a tilt toward higher
\( n \) and higher \( \omega _{b} \) gives therefore a better fit.
The effect is of the order of 10\% for \( k_{3}^{*}\approx 1 \). 

In the {}``gaussian skewness{}'' case the trend is qualitatively
the opposite, as can be seen in Fig. 3, where \( k_{3}^{*} \) ranges
from -4 to 4 (light to dark). Here the cosmological parameters decrease
for an increasing skewness. The reason is that now the effect is concentrated
around the intermediate multipoles \( \ell \approx 200 \): a positive
skewness induces smaller \( C_{\ell } \) at these multipoles, and
therefore a smaller \( n \) and \( \omega _{b} \) helps the fit.
The third case, the {}``hierarchical skewness{}'', is not shown
because is qualitatively similar to the previous case: the region
around \( \ell =200 \) is in fact also the region where \( C_{\ell } \)
is larger and therefore \( k_{3,\ell } \) given by Eq. (\ref{hier})
is larger.

\begin{figure}
{\centering \resizebox*{11cm}{!}{\includegraphics{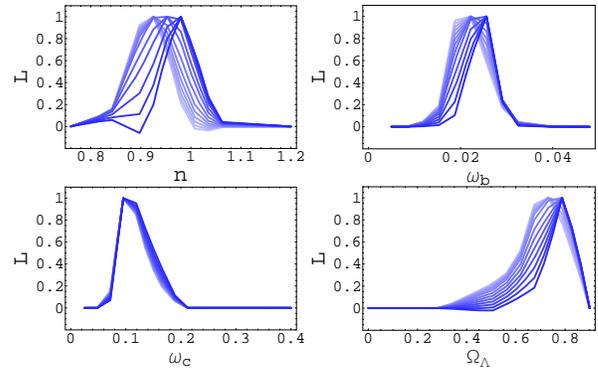}} \par}

\caption{Likelihood functions for the four cosmological parameters in the
constant skewness case. The skewness increases from -1.6 to 1.2, light
to dark.}
\end{figure}

\begin{figure}
{\centering \resizebox*{10cm}{!}{\includegraphics{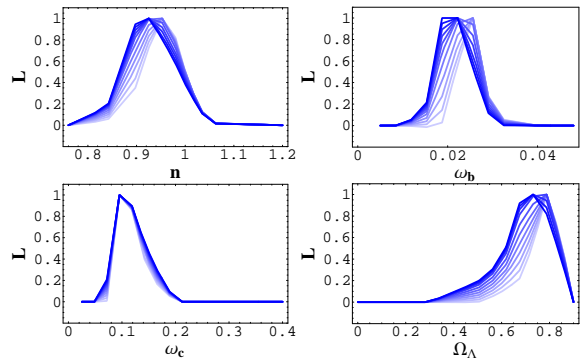}} \par}

\caption{Likelihood functions for the four cosmological parameters in the
{}``gaussian skewness{}'' case. The skewness increases from -4 to
4, light to dark.}
\end{figure}

Fig. 4 summarizes the results: the trend of the estimated parameters
(mean and standard deviation) versus \( k^{*}_{3} \) in the {}``constant
skewness{}'' case. The constant plateau that is reached for \( k_{3}^{*}<0 \)
depends on the fact that for large and negative skewness the peak
shift is independent of \( k_{3} \). The cosmological parameters
can be well fitted by the following expressions:

\begin{eqnarray}
n= & 0.90(1+0.03e^{k_{3}}), & \\
\omega _{b}= & 0.021(1+0.05e^{k_{3}}), & \\
\omega _{c}= & 0.115(1+0.025e^{k_{3}}), & \\
\Omega _{\Lambda }= & 0.67(1+0.06e^{k_{3}}). & 
\end{eqnarray}
For \( h \) the fit is \( h^{2}=0.42\frac{1+0.03e^{k_{3}}}{1-0.12e^{k_{3}}} \).
Notice that the trend for \( h \) is stronger than for the other
variables: \( h \) goes from 0.65 to 0.85 when \( k_{3} \) increases
from -1.6 to 1.2. Similar relations can be found for the other cases
as well.

\section{Conclusions}

This paper illustrates quantitatively a basic and obvious fact about
cosmological parameter estimation, namely the dependence on the underlying
statistics. Although the gaussianity of the CMB data is still to be
proved, almost all the previous works estimated the cosmological parameters
assuming vanishing higher order cumulants. Here it has been shown
that a non-zero skewness distorts the mode spectrum with respect to
the mean spectrum, inducing a considerable variation to the best fit
cosmological parameters. 

The Edgeworth expansion we used in this paper is convenient for analytical
purposes but its use is limited to relatively small deviations from
gaussianity. In fact, the peak shift displayed in Fig.1 is always
smaller than the errobars, and as a result the parameters, although
varying with \( k_{3} \), remain always within one sigma from the
zero-skewness case. This, however, does not mean that the dependence
on the higher order moments can be neglected, first because it is
a systematic effect, and second because more general probability distributions
which are not small deviations from gaussianity might introduce much
larger shifts.

We have shown that, to first order, the peak shift \( \Delta C_{\ell }/C_{\ell } \)
is proportional to \( k_{3,\ell }\sigma _{\ell } \). The error \( \sigma _{\ell } \)
includes cosmic variance and experimental errors. In the future, the
main source of error will be cosmic variance, at least below \( \ell =2000 \)
or so. A skewness of order unity will therefore introduce an additional
{}``skewness bias{}'' that will limit the knowledge of the cosmological
parameters by an amount similar to the cosmic variance itself. At
this point it will become necessary to estimate \( k_{3,\ell } \)
along with the other parameters. The first order EE is however inadequate,
since it is linear in \( k_{3,\ell } \), and it will be necessary
to extend the expansion to higher orders \cite{ame96, cont}, or to
adopt a non-perturbative non-gaussian distribution.

\begin{figure}
{\centering \resizebox*{9cm}{!}{\includegraphics{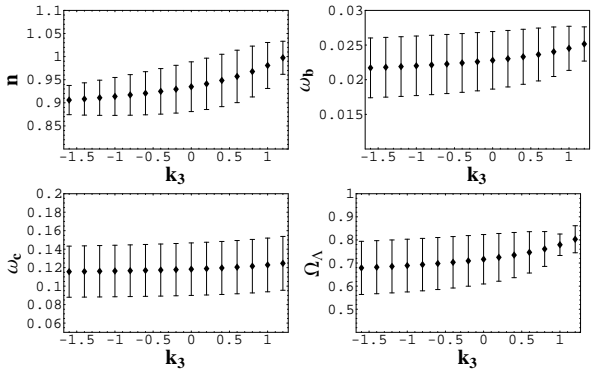}} \par}

\caption{Variation of the mean and standard deviation of the cosmological
parameters versus \protect\( k_{3}^{*}\protect \) in the {}``constant
skewness{}'' case. }
\end{figure}

\end{document}